\documentclass[sigconf]{acmart}
\def\BibTeX{{\rm B\kern-.05em{\sc i\kern-.025em b}\kern-.08emT\kern-.1667em\lower.7ex\hbox{E}\kern-.125emX}}
    
\usepackage{nicefrac}
\usepackage{siunitx}
\usepackage{array,framed}
\usepackage{adjustbox}
\usepackage{
  color,
  float,
  epsfig,
  wrapfig,
  graphics,
  graphicx,
  subcaption,
  adjustbox
}
\usepackage{textcomp}
\usepackage{setspace}
\usepackage{latexsym,fancyhdr,url}
\usepackage{enumerate}
\usepackage{algorithm2e}
\usepackage{algpseudocode}
\usepackage{graphics}
\usepackage{xparse} % argument parsing -- \edist
\usepackage{xspace}
\usepackage{multirow}
\usepackage{csvsimple}
\usepackage{balance}
\usepackage{
  tikz,
  pgfplots,
  pgfplotstable
}
\usepackage{hyperref}

\usetikzlibrary{
  shapes.geometric,
  arrows,
  external,
  pgfplots.groupplots,
  matrix
}

\usepackage{graphicx}
\usepackage{amsmath}
\usepackage{amsfonts}
\usepackage{bm}
\usepackage{multirow}
\usepackage{booktabs}
\usepackage{color}
\usepackage{array}
\usepackage{colortbl}
\usepackage{graphicx}
\usepackage{xcolor} 
\usepackage[normalem]{ulem}
\useunder{\uline}{\ul}{}

\usepackage{algpseudocode}
\usepackage{diagbox}
\usepackage[switch]{lineno}
\usepackage{makecell}
\usepackage{float}

\pgfplotsset{compat=1.9}
\usepackage{mathtools}

\DeclareMathAlphabet{\mathcal}{OMS}{cmsy}{m}{n}
\DeclareGraphicsExtensions{%
    .png,.PNG,%
    .pdf,.PDF,%
    .jpg,.mps,.jpeg,.jbig2,.jb2,.JPG,.JPEG,.JBIG2,.JB2}

\usepackage{xparse}
\newcommand{\bnm}{\begin{newmath}}
\newcommand{\enm}{\end{newmath}}

\newcommand{\bea}{\begin{eqnarray*}}%
\newcommand{\eea}{\end{eqnarray*}}%

\newcommand{\bne}{\begin{newequation}}
\newcommand{\ene}{\end{newequation}}

\newcommand{\bal}{\begin{newalign}}
\newcommand{\eal}{\end{newalign}}

\newenvironment{newalign}{\begin{align}%
\setlength{\abovedisplayskip}{4pt}%
\setlength{\belowdisplayskip}{4pt}%
\setlength{\abovedisplayshortskip}{6pt}%
\setlength{\belowdisplayshortskip}{6pt} }{\end{align}}

\newenvironment{newmath}{\begin{displaymath}%
\setlength{\abovedisplayskip}{4pt}%
\setlength{\belowdisplayskip}{4pt}%
\setlength{\abovedisplayshortskip}{6pt}%
\setlength{\belowdisplayshortskip}{6pt} }{\end{displaymath}}

\newenvironment{newequation}{\begin{equation}%
\setlength{\abovedisplayskip}{4pt}%
\setlength{\belowdisplayskip}{4pt}%
\setlength{\abovedisplayshortskip}{6pt}%
\setlength{\belowdisplayshortskip}{6pt} }{\end{equation}}

\newcounter{ctr}

\newcounter{mytable}
\def\mytable{\begin{centering}\refstepcounter{mytable}}
\def\endmytable{\end{centering}}

\newcounter{myfig}
\def\myfig{\begin{centering}\refstepcounter{myfig}}
\def\endmyfig{\end{centering}}

\newlength{\saveparindent}
\newlength{\saveparskip}
\newcommand{\E}{{\rm I\kern-.3em E}}

\renewcommand{\eqref}[1]{\mbox{Equation~(\ref{#1})}}

\def \part {part}

\DeclareMathOperator*{\argmin}{argmin}

\renewcommand{\paragraph}[1]{\vspace*{6pt}\noindent\textbf{#1}\;}

\def \blackslug{\hbox{\hskip 1pt \vrule width 4pt height 8pt
    depth 1.5pt \hskip 1pt}}
\def \qed{\quad\blackslug\lower 8.5pt\null\par}
\newcounter{mynote}[section]

\newcommand\ignore[1]{}

\newcounter{rcnote}[section]

\newcounter{mrnote}[section]

\newcounter{fknote}[section]

\newcounter{anote}[section]

\DeclareMathSymbol{\mlq}{\mathord}{operators}{``}
\DeclareMathSymbol{\mrq}{\mathord}{operators}{`'}

\newcommand{\rhf}[2]{R_{f, \gamma}}

\DeclareDocumentCommand{\edist}{o o}{
  \ensuremath{
    \IfNoValueTF{#1}{{d}}{{\sf d}(#1,#2)}
  }
}

\newcommand{\olrk}[1]{\ifx\nursymbol#1\else\!\!\mskip4.5mu plus 0.5mu\left(\mskip0.5mu plus0.5mu #1\mskip1.5mu plus0.5mu \right)\fi}

\NewDocumentCommand{\indseq}{ O{1} O{r} }{{#1}\ldots {#2}}

\begin{document}
%\fontfamily{lmr}\selectfont
% \def\thetitle{A Practical Way to Generate Strong Keys from Noisy Data}
\fancyhead{}
\def\thetitle{Generating Adversarial Point Clouds Using Diffusion Model}
\title{\thetitle}

\author[]{Ruiyang Zhao,Bingbing Zhu Chuxuan Tong, Xiaoyi Zhou, Xi Zheng}
% \author{Rahul Chatterjee}
% \affiliation{\small{UW--Madison}}

\date{}

\begin{abstract}
Adversarial attack methods for 3D point cloud classification reveal the vulnerabilities of point cloud recognition models. This vulnerability could lead to safety risks in critical applications that use deep learning models, such as autonomous vehicles. To uncover the deficiencies of these models, researchers can evaluate their security through adversarial attacks. However, most existing adversarial attack methods are based on white-box attacks. While these methods achieve high attack success rates and imperceptibility, their applicability in real-world scenarios is limited. Black-box attacks, which are more meaningful in real-world scenarios, often yield poor results. This paper proposes a novel black-box adversarial example generation method that utilizes a diffusion model to improve the attack success rate and imperceptibility in the black-box setting, without relying on the internal information of the point cloud classification model to generate adversarial samples. We use a 3D diffusion model to use the compressed features of the point cloud as prior knowledge to guide the reverse diffusion process to add adversarial points to clean examples. Subsequently, its reverse process is employed to transform the distribution of other categories into adversarial points, which are then added to the point cloud.
Furthermore, density-aware Chamfer distance is incorporated to constrain the noise added during back-propagation, further improving the imperceptibility of adversarial examples. Experimental results demonstrate that the proposed method exhibits high attack performance against various point cloud recognition models and defense methods, significantly enhancing the effectiveness of black-box attacks. In the black-box scenario, the attack success rate can reach about 90\%. The code for this work is available at: \url{https://github.com/AdvPC/Generating-Adversarial-Point-Clouds-Using-Diffusion-Model}.

% \jz{here you need to highlight the statistics of how much improved in terms of attack effectiveness compared with SOTA otherwise significant seems not supported}.

\end{abstract}
\maketitle

\section{Introduction}
%background、
% \cite{xiang2019generating}
% \IEEEPARstart{D}{eep} neural networks (DNNs) have contributed greatly to various computer vision tasks, including image classification and 3D point cloud recognition due to their human-exceeding performance \cite{fang2024explore}. However, recent studies have shown that DNNs are susceptible to adversarial examples (AEs) \cite{goodfellow, szegedy}. AEs can fool target models with human imperceptible perturbations. For example, researchers have implemented methods for generating adversarial examples of DNNs in speech \cite{yuan2018commandersong}, text \cite{jin2020bert}, and image scenarios \cite{liu2022practical}. Additionally, 3D point clouds are widely used as a critical component in autonomous driving \cite{zhuang20243d} systems. Recent studies have demonstrated that point cloud recognition models are vulnerable to adversarial examples \cite{cao, dong, sun, sun1, liu2020adversarial, tu, zhang}, effectively testing the robustness of 3D point cloud recognition models \cite{chu2022tpc} has become a pressing issue for researchers to address.  However, most current work on generating 3D adversarial examples focuses on either the white-box case or the black-box transfer-based approach. In contrast, the more realistic black-box case produces significant point perturbations or poor attack performance. This paper proposes a black-box adversarial examples generation method based on the diffusion model.

Deep neural networks (DNNs) have achieved remarkable success in various computer vision tasks, particularly in processing and analyzing 2D \cite{reddy2024deep,mesdaghi2024finger} and 3D data \cite{li2024bmmw,manisali2024efficient,yatbaz2024run}. However, studies show that DNNs are vulnerable to adversarial examples (AEs): carefully crafted perturbations added to clean samples that can mislead models without being noticeable to humans \cite{goodfellow, szegedy, zhang2024curvature, chen2024local}. Specifically, for 2D data, such as 2D images, sight color modifications on clean samples can fool recognition models and even pose significant threats to real-world applications \cite{eykholt2018robust}.
% For example, Eykholt \textit{et al.} \cite{eykholt2018robust} created adversarial traffic signs by adding graffiti and successfully fooled the target model during real-world driving-by tests. 
Similar adversaries can also be used on 3D point clouds.
Recent research shifted focus to the recognition of 3D point clouds due to their increasing importance in multiple fields \cite{enad2024detecting}. For instance, autonomous driving systems rely on LiDAR sensors to perceive and map the environment, because point clouds offer more precise geometric and structural information than 2D images \cite{zhuang20243d, wijaya2024advanced}. However, point cloud models are also found to be susceptible to AEs \cite{cao, dong, sun, sun1, liu2020adversarial, tu, zhang20233d}. Existing methods dominantly use white-box adversaries to maximize attack success rates and imperceptibility, whereas black-box settings remain underexplored on point cloud. In this work, we propose a novel approach for generating point cloud AEs with diffusion models, aiming to enhance attack success rate and stealthiness under black-box settings.

Point clouds are sets of unordered points that describe the shape of objects \cite{wang2024sequential}, and recognition models make predictions based on the representation of point cloud \cite{golovinskiy2009shape}. Shape latent is the compression feature of 3D point cloud. Therefore, generated adversarial point clouds are expected to keep shape changes perceptually negligible to humans, but capable of misleading recognition models. Several works extended existing gradient-based and optimization-based methods to point clouds \cite{liu2019extending, liu2020adversarial,xiang2019generating}, including Fast Gradient Sign Method (FGSM) \cite{liu2019extending}, Projected Gradient Descent (PGD) \cite{liu2020adversarial}, and Carlini and Wagner (C\&W) attack \cite{xiang2019generating}. In these methods, the generated perturbations are adjustable hyperparameters typically measured by Hausdorff distance \cite{zhang2017efficient} and Chamfer distance \cite{nguyen2021point}. However, they face a trade-off problem between imperceptibility and attack success rate. Higher attack success rates require more perturbations, while strong perturbations reduce imperceptibility. Another line of work aimed to minimize perturbations by adding, dropping, or shifting existing points. The modifications are based on manually designed rules either in greedy ways \cite{yang2019adversarial} or using optimization strategies \cite{ma2020efficient}. These pioneering explorations on point clouds are predominantly in white-box settings, which are less applicable to real-world scenarios compared to black-box settings. Existing black-box attacks are primarily limited to query-based methods but have yet to achieve comparable attack success rates and stealthiness \cite{huang2022shape, wicker2019robustness, liu2022imperceptible, naderi2022model}. Given the shortcomings of current adversarial examples generation schemes in black-box scenarios, we rethink the black-box adversarial examples generation method from the perspective of generative models.

In this work, we propose a novel black-box adversarial examples generation method, which uses the reverse diffusion process to add adversarial noise to the clean point cloud to craft adversarial examples. We regard AE generation as a reverse diffusion process, where the distribution of other classes is transformed into an adversarial distribution in the reverse-diffusion process. In addition, to enable the reverse-diffusion process to generate point clouds with obvious shape meanings, we use shape potential as prior knowledge. In the sample generation setting, the adversarial noise comes from the prior knowledge, and we use the normalizing flow \cite{chen2016variational} to parameterize the prior knowledge and drive the model to obtain strong expressive power. To further improve the interference-free execution of adversarial examples, we introduce the density-aware chamfer distance \cite{wu2021density} to bind the noise added during the back-propagation process. To evaluate the effectiveness of the black-box adversarial examples generation method using diffusion models, we evaluate our black-box adversarial examples generation method on common 3D point cloud recognition models, including PointNet2 \cite{qi2017pointnet++}, Curvenet \cite{muzahid2020curvenet}, PointConv \cite{wu2019pointconv} and compare it with optimization-based methods \cite{xiang2019generating} and generation-based methods \cite{chen2023diffusion}. The results demonstrate that our method successfully generates AEs capable of simultaneously fooling different point cloud recognition models. 
% Additionally, to evaluate the robustness of the adversarial example generation method, we evaluate the attack effectiveness against various defense methods\jz{this is also good}.\jz{what are the results?} We conducted attack experiments and comparative experiments to evaluate the effectiveness of our proposed AE generation method that does not rely on any 3D recognition model information\jz{what are the results?}.\jz{the sentences afterwards are too negative, you just need to summarize what are the future directions and you don't have to put negative results here. You can put detailed analysis in the results analysis section and you don't have to HIGHLIGHT NEGATIVE, THE POINT IS TO HIGHLIGHT THE FUTURE DIRECTIONS, THERE IS A HUGE DIFFERENCE} And rethink the work of 3D point cloud adversarial nature from a generative perspective. \textcolor{blue}{In addition, when facing complex surfaces, our black-box adversarial examples generation method does not improve significantly when facing other SOTA methods \cite{huang2022shape,lou2024hide}. Subsequently, we will consider combining the diffusion adversarial examples generation scheme with salient points for further research.}%We hope our work will inspire future research and applications in real-world 3D AI
The contribution of our work can be summarised as follows:
\begin{itemize}
    \item We propose a novel guided black-box adversarial method for generating adversarial point clouds. In our approach, point clouds from other classes are encoded into latent representations, which are conditions to guide the recovery process during reverse diffusion. The reverse diffusion process can automatically learn the significant features required to mislead the target model effectively under guidance.
    % We propose a black-box adversarial examples generation method that does not rely on 3D point cloud recognition model information.}
    % 
    
    \item We leveraged a novel loss function to generate adversarial point clouds with minimal perturbations. 
    % Based on the properties of the diffusion model, we propose a defense-resistant and well-transferable scheme.}
    
    \item The experimental data show that we can achieve over 90\% attack success rate when facing different point cloud recognition models even with defenses.
\end{itemize}

We will introduce this paper from the preliminaries, method, experiment results and conclusion.%preliminaries

\section{Related Work}
% \section{Related work}This chapter introduces the techniques used in this solution. We create adversarial samples and conduct experiments on the model introduced in Section 2.1. We use the defense method introduced in Section 2.2 for robustness testing.
We introduced a diffusion model to create adversarial samples and verified it in the commonly used 3D point cloud recognition model. To confirm the robustness of adversarial examples, we used several common defense methods for testing. Details are introduced in Sections 2.1-2.3.
\subsection{Point Cloud Recognition}
A point cloud is a sparse collection of points sampled by sensors to capture surface details \cite{guo2020deep}. Each point in the cloud specifies its position in the 3D coordinate system. Due to the unique characteristics of point clouds, models must learn representations from unordered inputs and extract information over both local and global geometric features. PointNet \cite{qi2017pointnet} is a benchmark model that directly takes raw point clouds as inputs and ensures permutation invariance through a symmetric function. PointNet++ \cite{qi2017pointnet++} incorporates a PointNet-based hierarchical structure to learn the neighborhood information of each point, further improving the recognition of local geometric information. Due to its simplicity and effectiveness, subsequent works have used PointNet++ as a backbone, extending it with attention mechanisms \cite{yang2019modeling, zhao2019pointweb, duan2019structural} or adaptive sampling strategies \cite{lin2019justlookup} for better local feature extraction and inference efficiency. Dynamic Graph CNN (DGCNN) \cite{wang2021object} builds on these ideas by dynamically updating the graph structure of the point cloud to capture local geometric relationships more effectively, enabling robust and flexible feature learning. CurveNet \cite{muzahid2020curvenet} further enhances this by representing the local geometry of points with learned curve segments, which improves the model's ability to capture fine-grained geometric details. PointConv \cite{wu2019pointconv} introduces convolution operations specifically designed for point clouds, enabling efficient and scalable learning of both local and global features. In this work, we leverage these advanced models to evaluate the performance and robustness of our proposed method for generating adversarial point clouds.
%cite{qi2017pointnet,te2018rgcnn,xu2021paconv,yang2018foldingnet,9577737,goyal2021revisiting,xiang2021walk}, has been widely used in the fields including 3D object categorization \cite{xiong2023enhancing}, scene segmentation \cite{rai2023unmasking}, and autopilot recognition \cite{cao}. PointNet \cite{qi2017pointnet} is a pioneering work in deep learning for 3D point clouds, which applies directly multilayer PointNet is a pioneering work in 3D point cloud deep learning, which directly applies multilayer perceptrons to learn point features and uses the max pool module to aggregate them in an efficient way. PointNet++ \cite{qi2017pointnet++} and many subsequent effective works \cite{duan2019structural,liu2019densepoint} are built on top of PointNet to further abstract the local features and combine them with the global features to obtain the information of the point cloud. 

% % 3D point cloud obtain, representaion (pc - triplets, mesh, etc)

\subsection{3D Adversarial Attacks and Defences}
% AEs can be crafted under either white-box or black-box settings \cite{}. \textcolor{blue}{a real-world example. => reduce the length}
% White-box adversaries have full knowledge of the target models, enabling them to generate powerful AEs like the Carlini \& Wagner (C\&W) attack \cite{} and the Jacobian-based Saliency Map Attack (JSMA) \cite{}. Conversely, black-box adversaries have limited access to the target model, typically restricted to model predictions. This limitation makes black-box attacks more challenging but also more applicable to real-world scenarios \cite{}. 

% the challenges in the AEs generation

Unlike image AEs directly adding perturbations, 3D AEs have to modify the shape of point clouds. However, large movements in shifting positions or changing the number of points bring unignorable perturbations to the perceptual quality. Therefore, current works focus on the trade-off between attack success rates and point movement. A few pioneering works extend gradient-based methods to 3D data that were originally developed for images. Liu \textit{et al.} \cite{liu2019extending} and Yang \textit{et al.} \cite{yang2019adversarial} applied the Fast Gradient Sign Method (FGSM) by constraining the movement of each point within a small range in $L^2$ norm. Further works \cite{liu2020adversarial, ma2020efficient, kim2021minimal, xiang2019generating, wen2020geometry, zhang20233d} improved Projected Gradient Descent (PGD) and Carlini and Wagner (C\&W) attacks by using either distance constraints or optimization functions to limit the point movement during AE generation. This enhances the smoothness of the adversarial point cloud, making the difference between it and the clean point cloud invisible to human eyes while achieving better attack performance. However, all of these methods require access to model-specific parameter details. Naderi \textit{et al.} \cite{wicker2019robustness} proposed a "model-free" approach that does not require knowledge of the target model. Tang \textit{et al.} \cite{Tang_Wu_Peng_Shi_Song_Gu_Tian_Wang_2023} use generative models for adversarial example generation, enabling the creation of high-quality and metastable adversarial examples without the need for model-specific information. The advantage of these black-box attacks lies in their ability to generate effective adversarial examples without prior knowledge of the target model's architecture or parameters, enhancing their applicability and versatility in various scenarios. The method we proposed is based on a generative diffusion model. Unlike previous generative approaches, we incorporate adversarial noise from the latent space perspective. This enhances the robustness, transferability, and quality of adversarial examples by mitigating outlier generation. 

% concealment ability = impercetability?

% Researchers have discovered that 3D neural networks can be misled by small perturbations applied to point clouds, making adversarial attacks a long-standing focus in the realm of deep learning. Current 3D adversarial attack methods have point generation attacks that add a limited number of points \cite{xiang2019generating} to the point cloud. Many researchers use gradient-based attacks to identify feature points from the point cloud for point removal attacks. There are also point perturbation attacks \cite{9156447,9577737, Hamdi2019AdvPCTA} based on Euclidean space, Among them, clicks are disturbed to deceive target classification through attack methods similar to 2D images, such as optimization problems (FGSM \cite{sen2023adversarial}, IFGSM \cite{you2023plant}, PGD \cite{villegas2024evaluating}). Evaluation metrics for point cloud attacks primarily focus on the effectiveness and imperceptibility of adversarial samples. Given the current challenges in achieving both strong imperceptibility and effective attack performance in black-box scenarios, this paper conducts an in-depth investigation into the issue using the diffusion model.

\subsection{Diffusion Models} 
In recent years, diffusion models have garnered significant attention in both academia and industry, demonstrating remarkable results across various applications. The diffusion process considered in this paper is closely related to probabilistic diffusion models, as referenced in \cite{DDPM,diffusion3D,dpm,nichol2022point}. Probabilistic diffusion models are a class of latent variable models that transform noise sampled from a Gaussian distribution into a data distribution using Markov chains. These models operate by iteratively adding and removing noise, effectively learning the underlying data distribution through a series of reversible transformations. While much of the existing work has focused on image-based diffusion models, our approach extends these concepts to the realm of 3D point clouds. In contrast to traditional image diffusion models, our diffusion model is specifically designed to handle 3D point cloud data by conditioning on latent variables to introduce noise into the 3D point cloud and subsequently rethinking point cloud black-box attacks from a latent variable perspective. By leveraging the inherent structure and properties of 3D point clouds, our method aims to enhance the robustness and accuracy of point cloud processing tasks. The application of diffusion models to 3D data presents unique challenges and opportunities, as the high-dimensional and unstructured nature of point clouds necessitates novel techniques for effective noise addition and removal. Our approach incorporates advanced probabilistic methods to manage these complexities, ensuring that the diffusion process preserves the geometric integrity and fine-grained details of the 3D structures. By conditioning on latent variables, our model introduces controlled perturbations to the point clouds, allowing for a nuanced examination of black-box attacks from a latent variable perspective. This enables us to develop more sophisticated and resilient defense mechanisms against such attacks, enhancing the overall robustness and accuracy of point cloud processing tasks. Our work demonstrates the potential of diffusion models to handle high-dimensional, unstructured data, paving the way for future research into leveraging these models for complex data structures and applications.

\section{Preliminaries}
% This work aims to generate highly imperceptible adversarial point clouds with transferability through the reverse process in 3D diffusion model. We first show problem formulation and overview of the framework, then present details of AE generation and loss functions below.

% \subsection{Problem Formulation}
% Given a clean point cloud $X$ and a well-trained target model $F_{\theta}$ for point cloud recognition, an adversary aims to fool the model $F_{\theta}$ by adding negligible perturbations $\delta$ to the clean sample: $F_{\theta}(X+\delta) \neq y$, where $\theta$ refers to the model parameters and $y$ is the correct label. 

% highly imperceptible: distance (evaluation)
% transferability? => attack other models 
% We focus on AE generation in the context of point cloud classification tasks. 
\subsection{Point Cloud}
Let $\{X,y\}$ be a point cloud and its corresponding label, where $X = \{x_i\}_{i=1}^{n}, x_i \in \mathcal{R}^3$ refers to $n$ points involved to represent a meaningful shape of a point cloud. Each single point $x_i$ has three dimensions to describe its location in the space. A 3D classifier $F$ learns spatial features of point clouds, which is tasked with correctly predicting the label of each: $F(X) = y$. In this work, we aim to mislead the classifier $F$ to output wrong predictions. 

% This work aims to generate adversarial PC by adding a small perturbation $\delta$: $X'=X+\delta$, misleading the classifier $F_\theta$ to output the wrong prediction:
% \begin{equation}
%     F_{\theta}(X +\delta) \neq y_i.
% \end{equation}

\subsection{Diffusion Probabilistic Model}
\label{pre:diffusion}
A standard probabilistic diffusion model on point clouds consists of two key processes \cite{DDPM,diffusion3D}: 1) a forward diffusion process that incrementally adds noise to the data, transforming it into pure noise, and 2) a reverse diffusion process that iteratively denoises the data to reconstruct the original input (e.g., an image or point cloud) \cite{DDPM,diffusion3D}. Assuming that each point $x_i$ in a point cloud is sampled independently from the same underlying distribution (e.g., a point distribution), we model the diffusion and reverse processes for individual points $x_i$.

The forward process gradually corrupts original data $x_i$ over a fixed number of steps $T$ \cite{DDPM,diffusion3D}:
\begin{equation}
q(x_{i}^{(1:T)}|x_{i}^{(0)})=\prod_{t=1}^{T} q(x_{i}^{(t)}|x_{i}^{(t-1)}),
\end{equation}
where $q(x_{i}^{(t)}|x_{i}^{(t-1)})$ refers to the noise-adding process at each step $t$ conditioned only on the previous step $t-1$ \cite{DDPM,diffusion3D}. Specifically, Gaussian noise is added iteratively as follows:
\begin{equation}
    q\left(\boldsymbol{x}^{(t)} \mid \boldsymbol{x}^{(t-1)}\right)=\mathcal{N}\left(\boldsymbol{x}^{(t)}; \sqrt{1-\beta_{t}} \boldsymbol{x}^{(t-1)}, \beta_{t} \boldsymbol{I}\right), t=1, \ldots, T,
\end{equation}
where $\beta_t\in(0,1)$ are hyperparameters controlling the noise level at each step. Larger values of $\beta$ bring more noise. As $t$ increases, the point cloud $x$ becomes increasingly noisy, eventually approaching pure Gaussian noise at step $T$.

% afterwards, in order to better represent the distribution of each point in the point cloud, we introduce the point cloud compression feature $z$ as a condition to join the diffusion process, which we express as $q(x_i^{(0)}|z)$. Here $z$ can be a point cloud compression feature different from the current classification.
The reverse process aims to recover meaningful point cloud structures from the noisy data generated by the forward process \cite{diffusion3D}. These meaningful structures are encoded into a latent representation $z$ which serves as a condition for reconstructing the original sample from step $T$ to $0$. The reverse process at each step is defined as
\begin{equation}
    p_{\boldsymbol{\theta}}\left(\boldsymbol{x}^{(0: T)} \mid \boldsymbol{z}\right) = p\left(\boldsymbol{x}^{(T)}\right) \prod_{t = 1}^{T} p_{\boldsymbol{\theta}}\left(\boldsymbol{x}^{(t-1)} \mid \boldsymbol{x}^{(t)}, \boldsymbol{z}\right), 
\end{equation}
\begin{equation}
p_{\boldsymbol{\theta}}\left(\boldsymbol{x}^{(t-1)} \mid \boldsymbol{x}^{(t)}, \boldsymbol{z}\right) = \mathcal{N}\left(\boldsymbol{x}^{(t-1)}; \boldsymbol{\mu}_{\boldsymbol{\theta}}\left(\boldsymbol{x}^{(t)}, t, \boldsymbol{z}\right), \beta_{t} \boldsymbol{I}\right),\,
\end{equation}
% We mainly generate adversarial point clouds using a back-diffusion process, which uses the compressed features $z$ of the point cloud to guide the model to add noise to the input examples. The reverse diffusion process can be defined as follows:
% \begin{multline}
% p_{\boldsymbol{\theta}}\left(\boldsymbol{x}^{(0: T)} \mid \boldsymbol{z}\right) = p\left(\boldsymbol{x}^{(T)}\right) \prod_{t = 1}^{T} p_{\boldsymbol{\theta}}\left(\boldsymbol{x}^{(t-1)} \mid \boldsymbol{x}^{(t)}, \boldsymbol{z}\right), \\
% p_{\boldsymbol{\theta}}\left(\boldsymbol{x}^{(t-1)} \mid \boldsymbol{x}^{(t)}, \boldsymbol{z}\right) = \mathcal{N}\left(\boldsymbol{x}^{(t-1)} \mid \boldsymbol{\mu}_{\boldsymbol{\theta}}\left(\boldsymbol{x}^{(t)}, t, \boldsymbol{z}\right), \beta_{t} \boldsymbol{I}\right),\,
% \end{multline}
where $\mu_\theta$ is an estimated mean through a neural network based on $x^{(t)}$ and the latent representation $z$ of the desired point cloud. Unlike prior diffusion-based methods for point cloud generation \cite{diffusion3D}, which rely on clean point clouds as guidance, our approach aims to generate adversarial examples. To achieve this, we condition the reverse process on latent representations $z$ associated with adversarial classes rather than clean samples. This adaptation allows the reverse diffusion process to generate adversarial point clouds with minimal perceptual noises.

\subsection{Problem Analysis}
The adversarial perturbation is added by the reverse diffusion process of the diffusion model. Since each reverse diffusion generation of the diffusion model can generate samples similar to the previous step, based on this feature, we assume that the reverse diffusion process \cite{diffusion3D} can generate an adversarial point cloud close to the previous step's sample point cloud. After iterating this process, we generate the adversarial examples we need. In order to limit the added perturbation $\delta$ from being too large, resulting in unexpected deformation of the generated adversarial examples, we use the density-attracted chamfer distance (DCD) to ensure the consistency of the point cloud, which enhances the robustness to local details and is computationally efficient. To further improve the concealment of point clouds, we introduce Mean Square Error(MSE). At the same time, this approach can prevent the original shape of the point cloud from being destroyed due to the generation diversity of the generation model itself. Subsequently, in order to improve the effectiveness of the attack, we need a query update module to improve the adversarial nature of the noise points and ensure that they effectively mislead the recognition model. Based on this idea, we propose an adversarial examples generation scheme based on the diffusion model, and the specific approach will be explained in detail in the next section.
% only point out one side of adversarial generation: imperceptibility - because of the back diffusion so that the AE could be close enough?
% lack of the attack effectiveness

% list formula of diffusion for later use?

\begin{figure*}[t]
    \centering
    \includegraphics[width=0.8\linewidth]{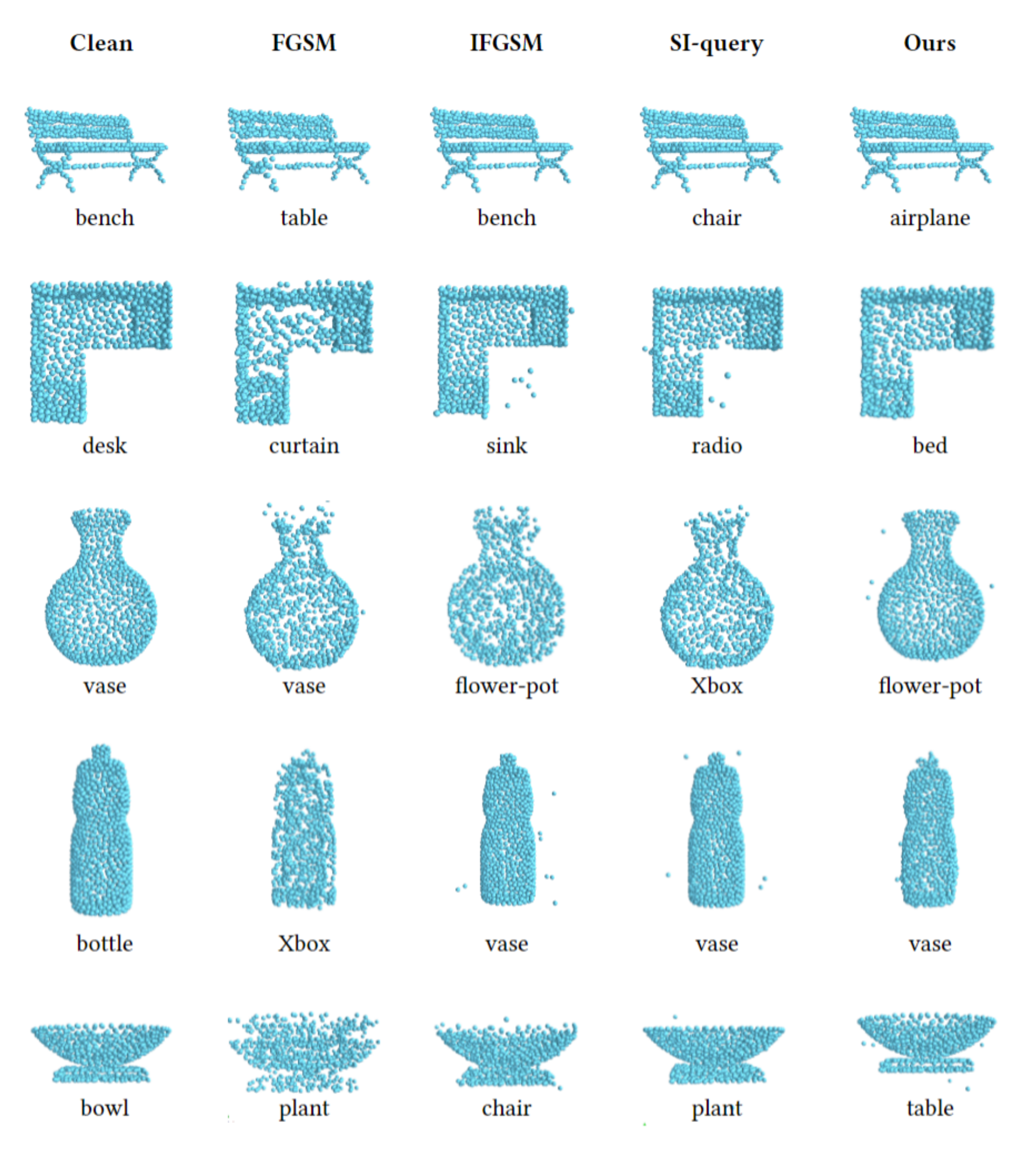}
    \caption{Comparison of various attack methods..}
    % Latent representation $z$ is calculated from the training phase of the 3D diffusion model \cite{diffusion3D}. $z$ guides noise generation and adversary use queries to increase the success rate of supply.}
    \label{fig:overview}
\end{figure*}

\section{Method}
% We first introduce the threat model, then present the implementation in the following as shown in Fig. \ref{fig:overview}.
This section outlines three steps for generating adversarial point clouds, as presented in Figure \ref{fig:overview}: 1) calculate the latent representations of guidance point clouds, 2) generate adversarial point clouds through a reverse diffusion process, 3) suppress the diversity of the diffusion model. We begin by introducing the threat model considered in this work.

% We aim to generate adversarial point clouds by adding perturbations extracted from the latent space of a 3D-diffusion model, then leverage the transferability of generated AEs to fool target model $F_{\theta}$. To enhance the imperceptibility of generated AEs, we constrain the perturbations with Mean Square Error (MSE) and Density-based Chamfer Distance (DCD) \cite{wu2021density} simultaneously. We present the details of AE generation below. The framework overview is shown in Figure \ref{fig:overview}.\jz{I think we need to start with an attack model as it is an attack paper. Also I suggest to walk through the main figure here in high-level, for instance, highlight in the figure what are the novelty part of our design, making them I, II, III and then delve-deeper into each of them, explaining the novelty in each of these components comparing with the SOTA}

\begin{figure*}[!t]
    \centering
    \includegraphics[width=\linewidth]{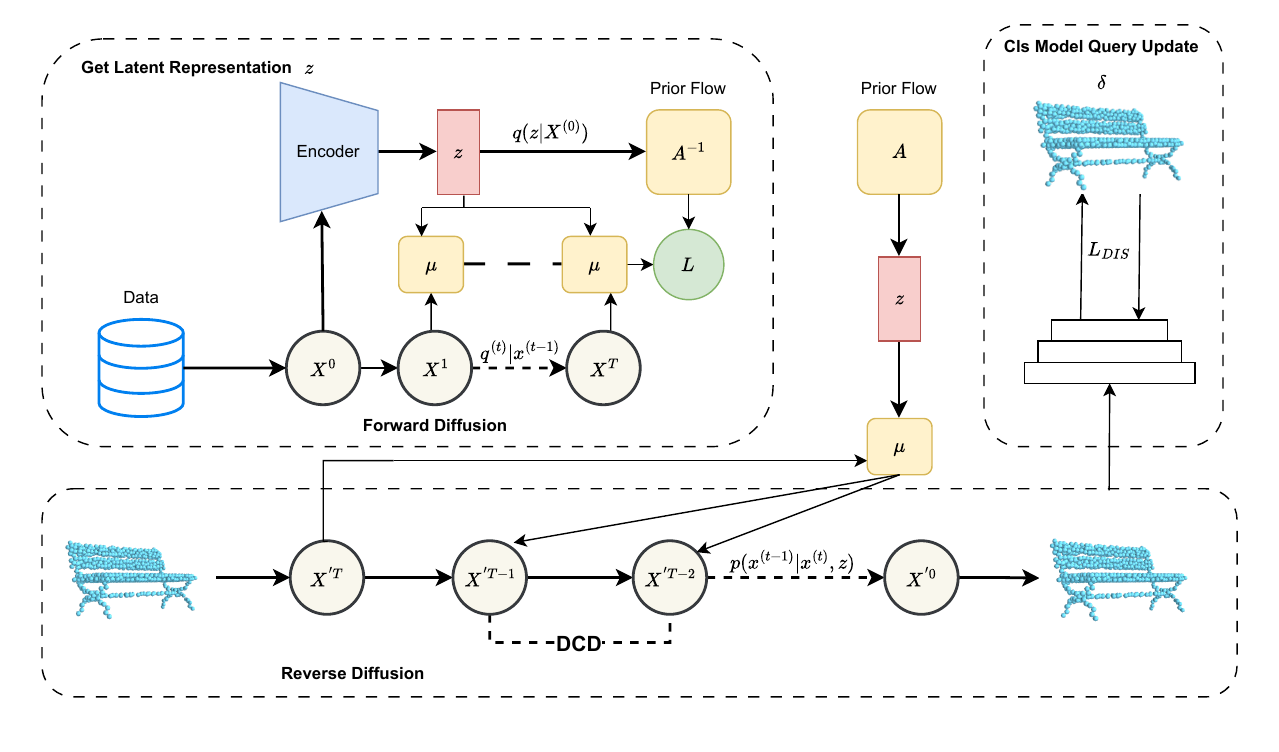}
    \caption{The overview of the proposed method.}
    % Latent representation $z$ is calculated from the training phase of the 3D diffusion model \cite{diffusion3D}. $z$ guides noise generation and adversary use queries to increase the success rate of supply.}
    \label{fig:overview}
\end{figure*}

\subsection{Threat Model}
% This paper focuses on black-box adversarial sample generation for the 3D point cloud classifier DNN. Specifically, the adversary intends to generate effective adversarial 3D point clouds without access to the specific details of the victim model to be deceived. The generated adversarial samples should be able to deceive other victim models to a certain extent. This attack may occur in real scenarios where the target model is deployed and the adversary cannot query the model.
This work focuses on generating adversarial point clouds under black-box settings to simulate real-world scenarios. The goal is to craft effective adversarial point clouds that mislead a target point cloud classifier $F$ by adding imperceptible noise $\delta$: 
\begin{equation}
    F(X + \delta) \neq y.
\end{equation}
In this setting, the adversary has limited access to the internal details of the target model $F$, such as its parameters or outputs. Additionally, the generated adversarial samples are designed to be transferable, enabling them to deceive unseen classifiers as well.

% Due to the generative nature of the model, which can produce diverse outputs, we apply constraints to ensure that the added noise remains effective and imperceptible: specifically, we use DCD loss to maintain the geometrical proximity between the transformed and original point clouds, and employ an MSE loss to ensure that the final distribution of the adversarial point cloud does not deviate significantly from the original, thus ensuring that the generated adversarial point clouds are both effective in fooling 3D models and maintain a high degree of similarity to the original point clouds, enhancing their stealthiness and transferability across different models.

% \subsection{Problem Formulation}
% Given a clean point cloud $X$ and a well-trained target model $F_{\theta}$ for point cloud recognition, an adversary aims to fool the model $F_{\theta}$ by adding negligible perturbations $\delta$ to the clean sample: $F_{\theta}(X+\delta) \neq y$, where $\theta$ refers to the model parameters and $y$ is the correct label. 

\subsection{Calculate latent representation $z$ as guidance}
Adversarial point clouds are meaningful structures that closely resemble the original input but include subtle perturbations. In standard diffusion models \cite{diffusion3D}, the meaningful structure of the desired shape is encoded and represented by latent representations $z$, which are used during the reverse diffusion process for recovery. $z$ is usually learned through a bottleneck layer from a variational autoencoder (VAE). However, unlike standard models, our task focuses on generating adversarial examples. Consequently, the latent representations $z$ are extracted from point clouds belonging to different classes (i.e., classes that do not overlap with the original class of the adversarial examples).

In this step, latent representation refers to the encoded abstract features of input data in a lower-dimensional space, capturing its essential structure for generative tasks. As the diffusion model selects points from the Gaussian distribution, we use a transformer-based model $A$ \cite{vaswani2017attention} to fuse the point cloud latent representation with the Gaussian noise to ensure that the added noise points are the adversarial points we need \cite{diffusion3D}, It can be expressed as:
\begin{equation}
    p(z)=p_{w}\left(A_{\alpha}^{-1}(z)\right) \cdot \frac{1}{\left|\operatorname{det} J_{A_{\alpha}}\left(A_{\alpha}^{-1}(z)\right)\right|}
    \label{A}
\end{equation}
where \(p_{w}\) comes from Gaussian distribution, \( F_\alpha^{-1}(z) \) represents the transformation from \( z \) through \( F_\alpha^{-1} \) back to \( w \), i.e. the transformation between the two. \( J_{F_\alpha}(w) \) is the Jacobian matrix of \( F_\alpha \). 

\( z \) represents latent representation, follows a conditional Gaussian distribution \( q(z|x^{(0)}) \), where \( \mu \) serves as the mean of the distribution, encapsulating the global structure of the input point cloud. $q^{(t)|x^{(t-1)}}$ is the forward diffusion step, as described in Section \ref{pre:diffusion}. We use forward diffusion to learn the latent representation of the point cloud as prior knowledge. $\mathcal{L}$ denotes the loss function \cite{diffusion3D}. As each reverse diffusion step generates a point cloud close to the previous step, we exploit this feature and use the adversarial latent representation $z$ of the current input to guide the AE generation process.

\subsection{Reverse Diffusion for Point Clouds}
% After training and obtaining the VAE’s latent representation, we use reverse diffusion to add noise to the point cloud. For the adversarial addition process of the clean point cloud, We introduce the reverse diffusion process into adversarial point cloud generation to produce transferable adversarial samples. At each step of the reverse diffusion process, appropriate noise is added based on the current point cloud features, generating point clouds highly similar to the original distribution $z$. By iteratively applying this process, the diffusion model can generate realistic samples. Utilizing this property, we use clean point clouds as the input for reverse diffusion and guide the process using data distribution \( z \) that does not belong to the current point cloud classification. This \( z \) is obtained as a prior distribution through training a self-supervised model with VAE, supervising the reverse diffusion process. This ensures that each step of diffusion, represented as \( p(x_i^T) \), includes adversarial points sampled from distributions other than the approximate input points. This method allows us to add noise to clean point clouds, transforming the reverse diffusion process into an adversarial noise addition process.
In this step, we denote the process of transforming a clean point cloud into an adversarial point cloud as the reverse diffusion process. The process can be defined as follows:
\begin{equation}
     x^{'}=Attack(x)=reverse(x^T) \cdots reverse(x^0),
\end{equation}
\begin{equation}
    \argmin \mathcal{L}_{Attack}= \mathcal{L}_{DIS}(x',x),
\end{equation}
where $x$ is the input clean point cloud, $Attack(\cdot)$ denotes the reverse diffusion operation, $x'$ is the adversarial examples, and reverse represents the diffusion of each time step $t$. We use $\mathcal{L}_{DIS}$ to conduct query attacks to improve the attack effectiveness of adversarial points in the reverse diffusion process, which will be introduced in the next step.
% The following formula can approximate the specific guidance method:
% \begin{multline}
% p_{\boldsymbol{\theta}}\left(\boldsymbol{x}^{(0: T)} \mid \boldsymbol{z}\right) = p\left(\boldsymbol{x}^{(T)}\right) \prod_{t = 1}^{T} p_{\boldsymbol{\theta}}\left(\boldsymbol{x}^{(t-1)} \mid \boldsymbol{x}^{(t)}, \boldsymbol{z}\right), \\
% p_{\boldsymbol{\theta}}\left(\boldsymbol{x}^{(t-1)} \mid \boldsymbol{x}^{(t)}, \boldsymbol{z}\right) = \mathcal{N}\left(\boldsymbol{x}^{(t-1)} \mid \boldsymbol{\mu}_{\boldsymbol{\theta}}\left(\boldsymbol{x}^{(t)}, t, \boldsymbol{z}\right), \beta_{t} \boldsymbol{I}\right),\,
% \end{multline}
% where $\mu_\theta$ is the mean value implemented by a neural network parameterized by $\theta$, the network is set as PointNet. $z$ is the shape latent control the noise. The starting $p(x_i^{T})$ is set to a clean input $X$. We obtain the adversarial examples with the latent shape $z$ by passing a set of points sampled from $p(x_i^{T})$ through the reverse process.
\subsection{Suppressing the Diversity of Diffusion Model}
To prevent the initial point cloud from losing its original shape due to the strong recovery capability of the reverse diffusion process to avoid the generation of excessive outlier points, and improve the concealment of adversarial point clouds we employ a Density-aware Chamfer Distance (DCD) optimization that increases with each reverse diffusion step. We use DCD distance as the loss function to control the diffusion process to ensure consistency between each successive point cloud. We demonstrate the effect of this in the subsequent experiment result section. The Density-aware Chamfer Distance (DCD) is defined as follows:
\begin{multline}
\mathcal{L}_{DCD}\left(X, X^{'}\right)=\min\frac{1}{2}\left(\frac{1}{\left|X\right|} \sum_{x \in X}\left(1-\frac{1}{n_{\hat{y}}} e^{-\alpha|| x-\hat{y} \|_{2}}\right)\right.
\\ \left.+\frac{1}{\left|X^{'}\right|} \sum_{y \in X^{'}}\left(1-\frac{1}{n_{\hat{x}}} e^{-\alpha|| y-\hat{x} \|_{2}}\right)\right),
\end{multline}
This distance extracts global features in the first stage and introduces local features with rich geometric information in the second stage to realize density perception. The detailed method can be found in \cite{wu2021density}. We use the distance as the loss function and minimize this loss in each step of the $reverse(\cdot)$ operation acting on the diffusion process, where $\hat{y} = min_{y \in X^{'}}|| x-y \|_{2}$, $\hat{x} = min_{y \in X}|| y-x \|_{2}$, and $\alpha$ denotes a temperature scalar. Here $n_{\hat{y}} = |X_1^y|$, Each $y$ contributes $\left|-\frac{1}{n_{y}} \sum_{x \in X^{y}} e^{-\|x-y\|_{2}}\right| \in[0,1]$ to the overall distance metric before averaging. This integration ensures that the generated adversarial samples remain imperceptible while effectively deceiving the classification model, thus enhancing the robustness and effectiveness of the diffusion attack scheme.

In addition, we use the Mean Squared Error (MSE) loss function to remove outliers. During the diffusion process, we use DCD instead of MSE because DCD maintains spatial coherence and reduces excessive outliers by focusing on the local structure of the point cloud. For final optimization, MSE is used to ensure global alignment. This balanced approach leverages DCD for local integrity during diffusion and MSE for overall alignment, resulting in high-quality adversarial point clouds with preserved geometric features. We present the results with different loss settings in subsequent ablation experiments. The MSE is defined as follows:
\begin{equation}
    \mathcal{L}_{MSE}\left(X, X^{\prime}\right)=\min(\frac{1}{n} \sum_{i=1}^{n}\left\|X_{i}-X_{i}^{\prime}\right\|_{2}^{2}).
\end{equation}
 where \(X'\) denotes the generated adversarial point cloud. The \(n\) stands for the total number of points in the point cloud, and \(X_i\) and \(X'_i\) are the \(i\)-th points in the original and generated point clouds, respectively. The term \(\| X_i - X'_i \|_2^2\) represents the squared Euclidean distance between \(X_i\) and \(X'_i\), quantifying the error for each point pair. We minimize MSE to achieve our goal because we found through experiments that minimizing MSE has little impact on the attack's success rate, but it can improve the concealment of the point cloud. We combine $\mathcal{L}_{DCD}$ with $\mathcal{L}_{MSE}$ to get the final optimization goal, as shown below:
 \begin{equation}
\mathcal{L}_{DIS}=\lambda_{1}\mathcal{L}_{DCD}+\lambda_2\mathcal{L}_{MSE}
 \end{equation}

\section{Experiment}

In this section, we present the experiment settings, experimental results, and ablation studies. Before elaborating on experimental details, we first set up research questions.

\subsection{Research Questions}
In this work, we explored generating point cloud AEs with diffusion models. To examine the attack success rate, concealment, and transferability of our method. We evaluate the method with the following research questions (RQs).
\begin{itemize}
    \item \textit{RQ1:} Given the limited research on generating 3D adversarial examples, can we use diffusion models to generate effective adversarial examples across different datasets and models, Such as PointNet++ and the ModelNet40 dataset?

    \item \textit{RQ2:} Can the 3D diffusion scheme generate adversarial examples resistant to defense when faced with different models?

  \item \textit{RQ3:} Explore whether different diffusion steps and noise constraints affect the performance of generated adversarial examples?
\end{itemize}

\begin{figure}[!t]
    \centering
    \includegraphics[width=0.5\textwidth]{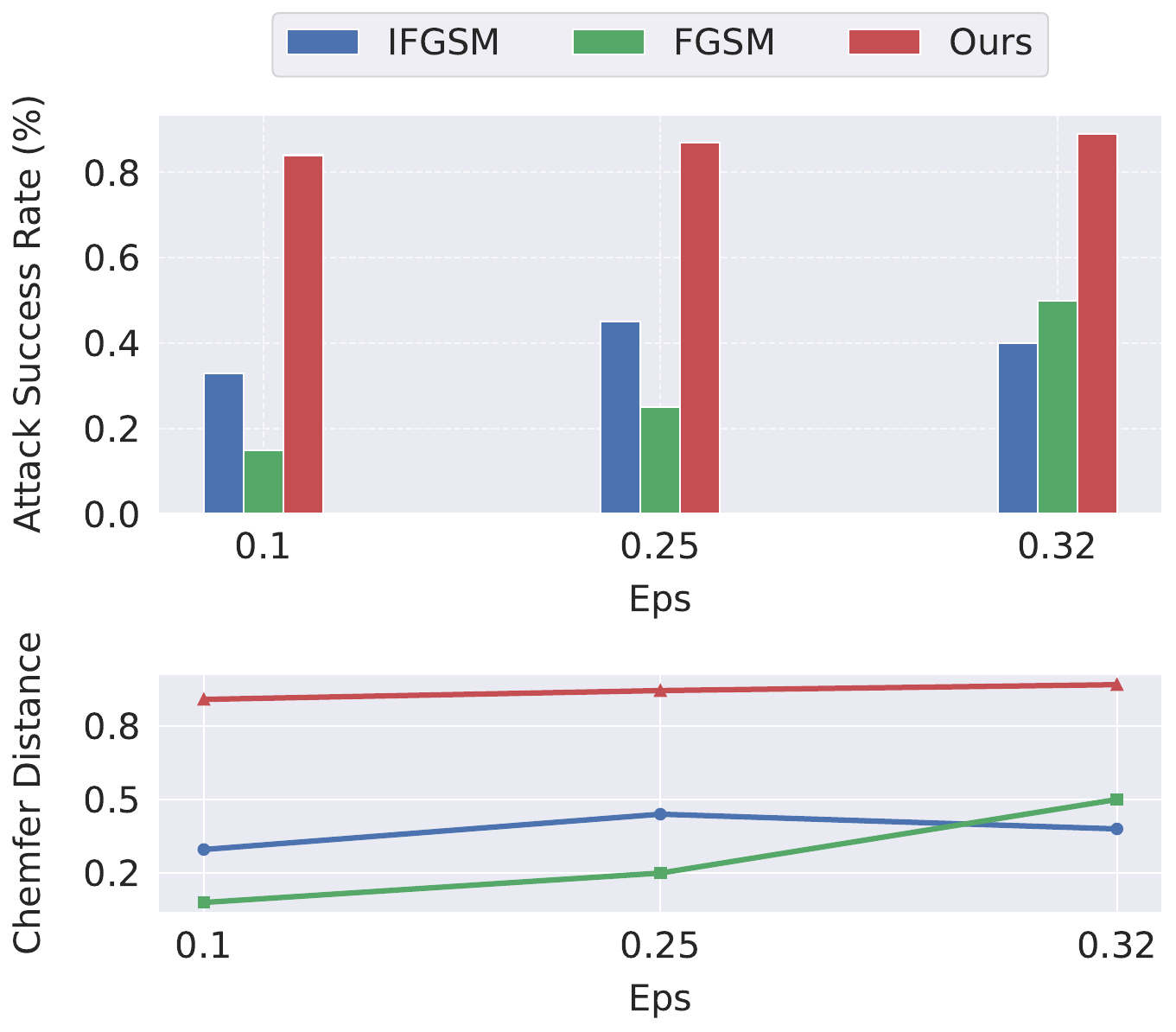}
    \caption{Effect of Eps on attack success rate (ASR) and Chamfer Distance(CD)}
    \label{fig:asr}
\end{figure}

\begin{figure}[!t]
\centering
\includegraphics[width=3in]{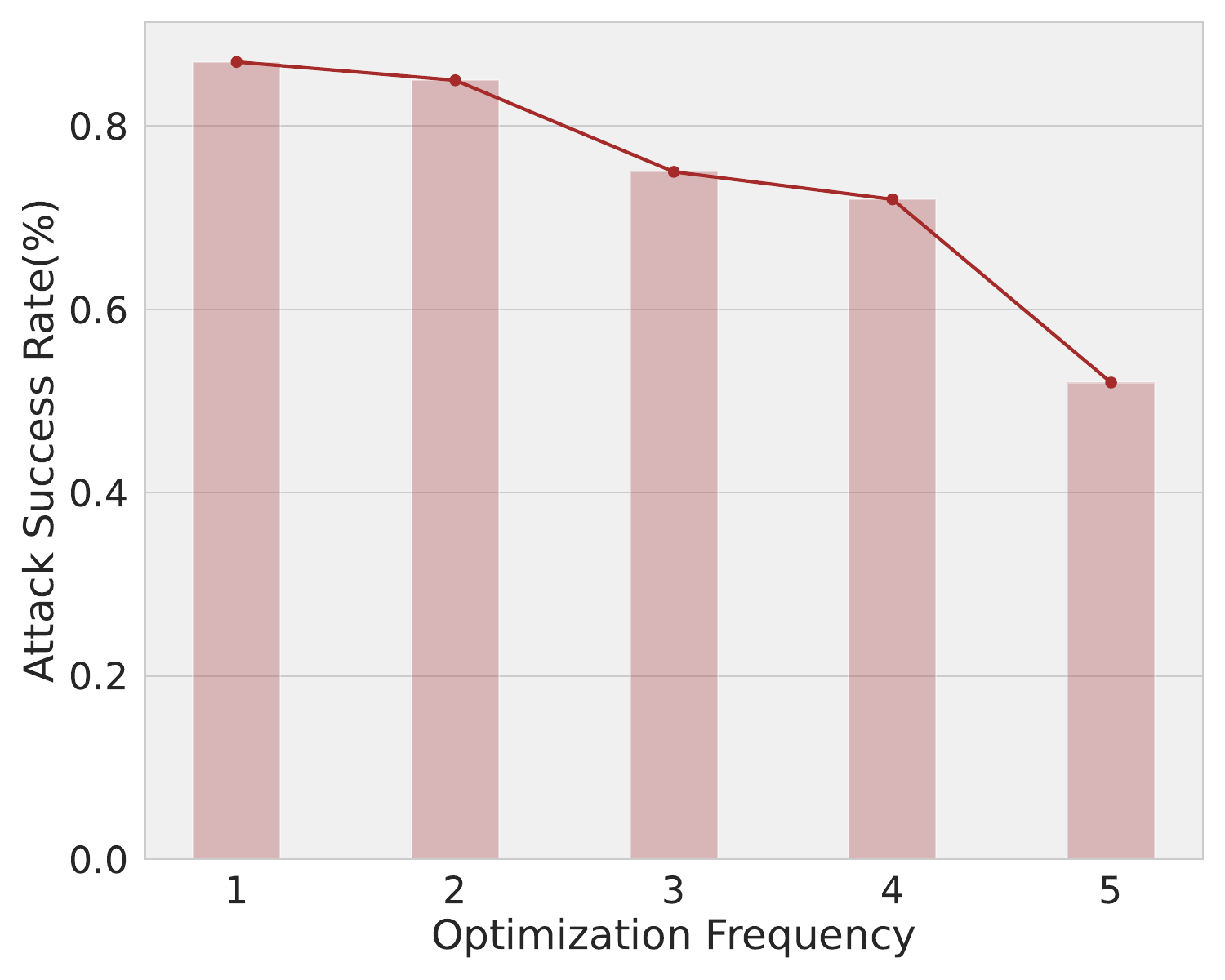}
\caption{Fix the level of noise added by diffusion at each step and optimize the effect of the number of times on the performance of the attack, targeting PointNet++.}
\label{fig:experiment_OP_ASR}
\end{figure}

\begin{table*}[!h]
\footnotesize
\centering
\caption{ASR (\%) of different attack methods with and without defense on  ModelNet40 (MN).}
    \begin{tabular}
    {lcp{10mm}<{\centering}p{10mm}<{\centering}p{10mm}<{\centering}p{10mm}<{\centering}p{10mm}<{\centering}p{10mm}<{\centering}p{10mm}<{\centering}p{10mm}<{\centering}p{10mm}<{\centering}p{10mm}<{\centering}p{10mm}<{\centering}p{10mm}<{\centering}p{10mm}<{\centering}}
        \toprule
        \multirow{3}{*}{Proxy Model}
        & \multirow{3}{*}{data}
        & \multirow{3}{*}{Defense}
        & \multicolumn{8}{c}{Attack Method} 
        \\
        \cmidrule(lr){4-12}
        & &  & \makecell{Drop\\-400} & \makecell{CW\\($l_2$)} & \makecell{CW\\(CD)} &\makecell{CW\\(HD)}& GeoA3 & AdvPC& \makecell{LG\\-GAN} & DPMA & Ours
        \\
        \midrule
        \multirow{3}{*}{PointNet} & \multirow{3}{*}{MN} & - & 59.64  & \textbf{100.00}  &  \textbf{100.00} & \textbf{100.00}  & \textbf{100.00}  & \textbf{100.00}  & 99.22  & 93.80  &  94.20
        \\
        &   & SRS & 58.14 & 53.00  & 67.19  & 69.94  & 81.65  & \textbf{98.87}  & 92.13  & 93.48 & 90.43    
        \\
         &   & SOR & 56.28 & 13.82  & 15.63   & 15.80  & 42.79  & 46.19  & 67.25 & 87.00  &   \textbf{93.06}
        \\
        \midrule
        \multirow{3}{*}{Dgcnn} & \multirow{3}{*}{MN} & - &  45.91 & \textbf{100.00}  & \textbf{100.00}  &  \textbf{100.00}   & \textbf{100.00} & 94.58  &  86.08 &97.45 & 95.23 
        \\
        &  & SRS &35.05   & 31.09  &  37.11 & 32.29  & 77.71  & 70.63  & 80.60  & \textbf{94.73}  & 92.40  
        \\
        &  & SOR &15.03   & 2.26  & 2.92  & 3.13  & 56.25  & 11.04  & 50.17  & 93.11  & \textbf{94.60}
        \\
        \midrule
        \multirow{3}{*}{PointConv} & \multirow{3}{*}{MN} & - & 37.12 &\textbf{100.00}  &  \textbf{100.00} & \textbf{100.00}  & 96.09 & 98.54  &  78.04 & 94.98   & 94.50
        \\
        &  & SRS  & 35.09  & 37.29  & 28.95  &  27.77 & 21.48  & 93.54  &  71.88 & \textbf{94.06} & 92.60
        \\
        & & SOR  & 34.44 & 18.13  & 17.29  & 19.16 &18.35  & 91.25 & 63.88  & 90.71  & \textbf{93.21}
        \\
        \bottomrule
    \end{tabular}
    % }
    % \vspace{0.5em}
    
    \label{tab:wh}
    % \vspace{-0.5em}
\end{table*}

\begin{table*}[!h]
    % \setlength{\belowcaptionskip}{1cm}
    % \scriptsizes
	\footnotesize
    \centering
        \caption{Quantitative comparison between our method and existing black-box transfer-based attacks in terms of attack success rate (ASR), Chamfer distance (CD), Hausdorff distance (HD), and the proxy model used, where CD is multiplied by $10^2$ and HD is multiplied by $10^2$ for better comparison.}
    \setlength{\tabcolsep}{1mm}{
    \begin{tabular}{lcp{10mm}<{\centering}p{10mm}<{\centering}p{10mm}<{\centering}p{10mm}<{\centering}p{10mm}<{\centering}p{10mm}<{\centering}p{10mm}<{\centering}p{10mm}<{\centering}p{10mm}<{\centering}p{10mm}<{\centering}p{10mm}<{\centering}p{10mm}<{\centering}}
        \toprule
        \multirow{3}{*}{Proxy Model}
        & \multirow{3}{*}{Attack}
        & \multicolumn{3}{c}{PointNet++ \cite{qi2017pointnet++}} 
        & \multicolumn{3}{c}{Curvenet \cite{muzahid2020curvenet}}
        & \multicolumn{3}{c}{PointConv \cite{wu2019pointconv}}
        \\
        \cmidrule(lr){3-5}\cmidrule(lr){6-8}\cmidrule(lr){9-11}
         & & ASR$\uparrow$ & CD$\downarrow$ & HD$\downarrow$ & ASR$\uparrow$ & CD$\downarrow$ & HD$\downarrow$ & ASR$\uparrow$ & CD$\downarrow$ & HD$\downarrow$  
        \\
        & & (\%) & ($10^{-2}$) & ($10^{-2}$) & (\%) & ($10^{-2}$) & ($10^{-2}$) & (\%) & ($10^{-2}$) & ($10^{-2}$) 
        \\
        \midrule
        \multirow{3}{*}{PointNet \cite{qi2017pointnet}} & FGSM \cite{sen2023adversarial}   & 68.2 & 2.93 & 12.3 &  75.7 & \textbf{0.60}  & \textbf{5.04}  & 71.9  & 2.18  & 17.06  
        \\
        & IFGSM \cite{you2023plant}  & 78.0 & 2.15  & 16.7  &  77.0 &  1.62 & 12.7  & 76.0 & \textbf{1.06}  & 11.8   
        \\
         & PGD \cite{villegas2024evaluating}  & 70.17  &  9.8 & 39.1  & 67.6  & 9.83  & 46.9  &  61.2 & 9.7  & 46.2  
        \\
        \midrule
        \multirow{3}{*}{Dgcnn \cite{wang2021object}} & FGSM \cite{sen2023adversarial}   & 41.3 & 2.52 & 9.13 & 50.2 & 2.52  & 9.13  & 55.2 & 4.09 & 14.29 
        \\
        & IFGSM \cite{you2023plant}  & 57.6 & 3.5 & 12.5 &  72.5 & 3.49  & 12.5  & 60.2 & 3.5  & 12.52  
        \\
        & PGD \cite{villegas2024evaluating}  & 61.6  & 9.63  & 38.1  & 62.5  & 9.73  & 46.9 & 62.8  & 9.8  & 48.3 
        \\
        \midrule
        \multirow{3}{*}{Pct \cite{guo2021pct}} & FGSM \cite{sen2023adversarial}  &  30 & 3.7  & 14  & 45.2  & 3.75  & 13.9  & 45.6  & 3.2  & 12.4
        \\
        & IFGSM \cite{you2023plant}  & 44.5  & 5.5  & 19.3  & 72.3  & 5.56  & 19.3 &  64.3 & 5.56  & 19.3
        \\
        & PGD \cite{villegas2024evaluating} & 69  & 9.9  & 48.3  &  63.4 & 9.9  & 48  & 63.0  & 9.8  & 48.3        \\
        \midrule
        3D-Diffusion& (DifA) Ours  & \textbf{86.0}  & \textbf{1.7}  & \textbf{7.4}  & \textbf{87}  &  1.7 & 7.4  & \textbf{86}  & 1.7  & \textbf{7.4} 
        \\
        \bottomrule
    \end{tabular}
    }
    % \vspace{0.5em}
    \label{tab:table1}
    % \vspace{-0.5em}
\end{table*}

\subsection{Experimental Settings }
\textbf{Datasets. }
For a fair comparison, we evaluate AEs generated by our method on ShapeNet and ModelNet40 as prior works \cite{huang2022shape}. We assess the attack performance on ShapeNet, which includes approximately 50,000 3D CAD models across 14 major and 55 subcategories, with each model containing at least 2,000 points \cite{chang2015shapenet}. Additionally, we evaluate our approach on ModelNet40, consisting of 12,311 CAD models from 40 object categories, with 9,843 for training and 2,468 for testing \cite{sun2022modelnet40}. Each object in ModelNet40 is uniformly sampled to 2,048 points and rescaled to a unit cube. Data augmentation techniques, such as random scaling and jittering, are applied to preprocess the point clouds in the test set.

\textbf{RQ1: } To test whether our proposed method can generate effective adversarial point clouds on different datasets, we tested it on both ModelNet40 and ShapeNet datasets. We compare the attack success rate with four different SOTA and other optimization-based methods and conduct experimental comparisons on three different recognition models. The test results of ModelNet40 can be found in Table \ref{tab:wh}, and the test results of ShapeNet dataset are shown in Table \ref{tab:table4}.

 \textbf{RQ2: }We report the attack access rate with non-defense on ModelNet40 and ShapeNet. While ModelNet40 shows more distinguishable results than ShapeNet, we report the results of the rest tasks on ModelNet40. For ModelNet40, we chose three surrogate models for black-box transfer-based attack evaluation and compared them with our proposed approach. For ShapeNet, we use the proposed method to evaluate the attack performance of two models.

\noindent \textbf{Models. } 
Our approach is evaluated on six benchmark 3D recognition models: PointNet \cite{qi2017pointnet}, PointNet++(MSG) \cite{qi2017pointnet++}, PointConv \cite{wu2019pointconv}, Pct \cite{guo2021pct}, DGCNN \cite{wang2021object}, and Curvenet \cite{muzahid2020curvenet}. These models were chosen for their unique architectures and established performance in 3D recognition tasks:
\begin{enumerate}
    \item PointNet uses a symmetric function for permutation invariance, while PointNet++ introduces hierarchical feature learning. 
    \item PointConv enhances feature representation with a point cloud-specific convolution, and Pct excels with a transformer-based architecture. 
    \item DGCNN captures local structures with dynamic graph construction.
    \item Curvenet uses 3D curves for feature extraction. 
\end{enumerate}
We selected PointNet, DGCNN, and Pct as proxy models to conduct comparative tests on PointNet++, Curvenet, and PointConv, evaluating our approach's robustness with and without defense measures. This evaluation across diverse architectures allows us to assess the generalizability and effectiveness of our method, providing insights into its performance against various 3D recognition systems.

\noindent \textbf{Baselines. }  
We leverage a pre-trained open-source 3D-diffusion model, trained on extensive point cloud data, as the basis for our manifold attack and compare our approach (Ours) with eight baseline methods.  These include the deletion-based method Drop-400 \cite{zheng2019pointcloud}, which drops the most critical 400 points, and perturbation-based methods using optimization like C$\&$W under \( l_2 \)-norm, Chamfer distance (CD), and Hausdorff distance (HD) constraints \cite{xiang2019generating}. Additionally, we consider GeoA3 \cite{wen2020geometry}, which applies geometric-aware constraints, and AdvPC \cite{hamdi2020advpc}, which focuses on high transferability. We also compare against generative-based methods such as LG-GAN \cite{zhou2020lg} and DPMA \cite{tang2023deep}.
For attack performance testing, we select the best configuration of these adversarial attack methods to achieve the best attack success rate\cite{tang2023deep} they can achieve. Due to time constraints, we selected only FGSM, IFGSM, and PGD as comparative experiments for evaluating stealthiness.

\noindent \textbf{Evaluation Metrics. }
We begin by statistically analyzing the attack success rate (ASR) of the generated adversarial samples on the 3D point cloud recognition model to evaluate the method's effectiveness. To assess the imperceptibility and efficacy of the generated adversarial point clouds, we compute the Hausdorff Distance and Chamfer Distance between the original point cloud and the adversarial output. The Hausdorff Distance quantifies the maximum deviation between point sets, while the Chamfer Distance measures the average displacement, providing insights into the subtlety of the perturbations. The formula is defined as follows.
\begin{equation}
    \text{ASR} = \frac{\sum_{i=1}^{N} \mathbb{I}(Y_i \neq \hat{Y}_i)}{N} \times 100\% ,
\label{ASR}
\end{equation}
\begin{equation}
    \mathcal{D}_{C}\left(X, X^{\prime}\right)=\frac{1}{\left\|X^{\prime}\right\|_{0}} \sum_{y \in X^{\prime}} \min _{x \in X}\|x-y\|_{2}^{2}
    \label{CD}
\end{equation}
\begin{equation}
    \mathcal{D}_{H}\left(X, X^{\prime}\right)=\max _{y \in X^{\prime}} \min _{x \in X}\|x-y\|_{2}^{2},
    \label{HD}
\end{equation}
In these formulas, \(X\) represents the original point cloud, and \(X'\) denotes the adversarial point cloud. Equation \ref{ASR} defines the Attack Success Rate (ASR), which quantifies the percentage of perturbed samples misclassified by the model. \(Y_i\) and \(\hat{Y}_i\) represent the true and predicted labels, respectively, and \(N\) is the total number of samples. Equation \ref{CD} defines the Chamfer Distance (\(\mathcal{D}_C\)), which calculates the average displacement between points in \(X\) and \(X'\). Equation \ref{HD} defines the Hausdorff Distance (\(\mathcal{D}_H\)), which measures the maximum deviation between points in the two sets. The ASR, defined in Equation \ref{ASR}, quantifies the ratio of perturbed examples incorrectly classified by the black-box transfer-based attacked model. Evaluating our attack in terms of effectiveness, stealthiness, and transferability offers a comprehensive understanding of its impact on 3D recognition models.

\begin{itemize}
    \item \emph{Effectiveness}: Measured by the attack success rate (ASR), which reflects the percentage of adversarial examples that successfully mislead the target model. A higher ASR indicates a more effective attack.
    \item \emph{Stealthiness}: Assessed using the Hausdorff distance and Chamfer distance to ensure that the generated adversarial examples are imperceptible and maintain a high degree of similarity to the original point clouds.
    \item \emph{Transferability}: Evaluated by testing the adversarial examples on different models to determine the robustness and generalization of the attack across various 3D recognition systems.
\end{itemize}

% \textbf{RQ2: } To test whether our proposed method can generate effective adversarial point clouds on different datasets, we tested it on both ModelNet40 and ShapeNet datasets. The test results of ModelNet40 can be found in Tab \ref{tab:wh}, and the test results of ShapeNet dataset are shown in Tab \ref{tab:table4}.

\textbf{RQ3: }We use ablation experiments to find out the impact of different diffusion steps and noise constraints on the attack performance of generated adversarial examples. Details can be seen in Ablation Studies.

% \jz{this is odd, we need to explain RQ in order. RQ1 shall be across datasets, across pointcloud models and compared with Baselines. RQ2 shall be ablation study. RQ3 shall be test against defence. Reorganize the settings in order and in logical sequence}

\subsection{Experimental Results - RQ1}
\textbf{Performance Comparison with White Box Attacks and Generative Attacks. }
 The result in Table \ref{tab:wh} shows that Drop-400 performs the worst. Among them, the three CW attacks (using l2, Chamfer Distance, and Hausdorff Distance as loss functions), GeoA3, and AdvPC all showed a 100\% success rate on PointNet. In particular,  LG-GAN \cite{zhou2020lg}, DPMA \cite{tang2023deep} and our method also successfully attacked these models, but the success rate was slightly lower than the previous methods. This is because the previous methods are white-box attack methods, which can use the gradient information of the attacked model, which greatly improves the success rate of the attack. However, our proposed method differs from LGAN and DPMA in that our method does not use any information from the attacked model, and our method is dataset-oriented rather than sample-oriented. In addition, the attack success rate of AdvPC on Dgcnn and PointConv did not reach 100\%, due to the trade-off imposed by auto-encoder for transferability.

\begin{table*}[!h]
    % \setlength{\belowcaptionskip}{1cm}
    % \scriptsize
	\footnotesize
    \centering
        \caption{Quantitative comparison between our method and existing black-box transfer-based attacks in terms of attack success rate (ASR), Chamfer distance (CD), Hausdorff distance (HD), the proxy model and defense method SOR used, where CD is multiplied by $10^2$ and HD is multiplied by $10^2$ for better comparison.}
    \setlength{\tabcolsep}{1mm}{
    \begin{tabular}{lcp{10mm}<{\centering}p{10mm}<{\centering}p{10mm}<{\centering}p{10mm}<{\centering}p{10mm}<{\centering}p{10mm}<{\centering}p{10mm}<{\centering}p{10mm}<{\centering}p{10mm}<{\centering}p{10mm}<{\centering}p{10mm}<{\centering}p{10mm}<{\centering}p{10mm}<{\centering}}
        \toprule
        \multirow{3}{*}{Proxy Model}
        & \multirow{3}{*}{Attack}
        & \multirow{3}{*}{Defense}
        & \multicolumn{3}{c}{PointNet++ \cite{qi2017pointnet++}} 
        & \multicolumn{3}{c}{Curvenet \cite{muzahid2020curvenet}}
        & \multicolumn{3}{c}{PointConv \cite{wu2019pointconv}}
        \\
        \cmidrule(lr){4-6}\cmidrule(lr){7-9}\cmidrule(lr){10-12}
        & & & ASR$\uparrow$ & CD$\downarrow$ & HD$\downarrow$ & ASR$\uparrow$ & CD$\downarrow$ & HD$\downarrow$ & ASR$\uparrow$ & CD$\downarrow$ & HD$\downarrow$  
        \\
        & & & (\%) & ($10^{-2}$) & ($10^{-2}$) & (\%) & ($10^{-2}$) & ($10^{-2}$) & (\%) & ($10^{-2}$) & ($10^{-2}$) 
        \\
        \midrule
        \multirow{3}{*}{PointNet} & FGSM \cite{sen2023adversarial}  & \multirow{3}{*}{SOR} &  54.3 & \textbf{1.2}  & 16.3  & 66.5  & 1.6  & \textbf{2.05}  & 54.3  & 1.2  & 16.3  
        \\
        & IFGSM \cite{you2023plant} &  & 75.3  & 4.8  & 38.5  & 80  & 4.3  & 30.5  &  72 & \textbf{1.02}  & 12.2 
        \\
         & PGD \cite{villegas2024evaluating} &  & 69.5  & 9.9  & 48.2   & 82  & 9.95  & 48.2  & 80 & 9.7  & 46.4  
        \\
        \midrule
        \multirow{3}{*}{Dgcnn} & FGSM \cite{sen2023adversarial}  & \multirow{3}{*}{SOR} & 45.05 & 2.72  & 11.68  & 61  & 2.7  & 11.6  &  64.8 & 3.84  & 14.5  
        \\
        & IFGSM \cite{you2023plant} &  &  48.5 & 5.07 & 18.4  & 73.2  & 3.84 & 14.5  &  73 & 3.84   & 14.4   
        \\
        & PGD \cite{villegas2024evaluating} &  & 69.5  & 9.9  & 48.2  & 83  & 9.75  & 48.1  & 82  & 9.8  & 47.6 
        \\
        \midrule
        \multirow{3}{*}{Pct} & FGSM \cite{sen2023adversarial} & \multirow{3}{*}{SOR} & 46.8  & 2.75  & 11.68  & 61.8  & 2.7  &11.6   & 54.9  & 2.7  & 11.6
        \\
        & IFGSM \cite{you2023plant} &   & 47.8  & 5.09  & 18.5  &  77.8 & 5.09  & 18.5  & 65.6  & 5.06  & 18.4
        \\
        & PGD \cite{villegas2024evaluating}&   & 72.7  & 9.8  & 48.3  & 72.5  & 9.8  & 48.3  &  81 & 9.8  & 48.3
        \\
        \midrule
        3D-Diffusion& (DifA)Ours & SOR & \textbf{84.7}  & 1.4  & \textbf{6.8}  & \textbf{86.2}  & \textbf{1.4}  & 6.8  & \textbf{85.0}  & 1.4  & \textbf{6.8} 
        \\
        \bottomrule
    \end{tabular}
    }
    % \vspace{0.5em}
    \label{tab:table2}
    % \vspace{-0.5em}
\end{table*}

\begin{table*}[!htb]
    % \setlength{\belowcaptionskip}{1cm}
    % \scriptsize
	\footnotesize
    \centering
        \caption{Quantitative comparison between our method and existing black-box transfer-based attacks in terms of attack success rate (ASR), Chamfer distance (CD), Hausdorff distance (HD), the proxy model and defense method SRS used, where CD is multiplied by $10^2$ and HD is multiplied by $10^2$ for better comparison.}
    \setlength{\tabcolsep}{1mm}{
    \begin{tabular}{lcp{10mm}<{\centering}p{10mm}<{\centering}p{10mm}<{\centering}p{10mm}<{\centering}p{10mm}<{\centering}p{10mm}<{\centering}p{10mm}<{\centering}p{10mm}<{\centering}p{10mm}<{\centering}p{10mm}<{\centering}p{10mm}<{\centering}p{10mm}<{\centering}p{10mm}<{\centering}}
        \toprule
        \multirow{3}{*}{Proxy Model}
        & \multirow{3}{*}{Attack}
        & \multirow{3}{*}{Defense}
        & \multicolumn{3}{c}{PointNet++ \cite{qi2017pointnet++}} 
        & \multicolumn{3}{c}{Curvenet \cite{muzahid2020curvenet}}
        & \multicolumn{3}{c}{PointConv \cite{wu2021density}}
        \\
        \cmidrule(lr){4-6}\cmidrule(lr){7-9}\cmidrule(lr){10-12}
        & & & ASR$\uparrow$ & CD$\downarrow$ & HD$\downarrow$ & ASR$\uparrow$ & CD$\downarrow$ & HD$\downarrow$ & ASR$\uparrow$ & CD$\downarrow$ & HD$\downarrow$  
        \\
        & & & (\%) & ($10^{-2}$) & ($10^{-2}$) & (\%) & ($10^{-2}$) & ($10^{-2}$) & (\%) & ($10^{-2}$) & ($10^{-2}$) 
        \\
        \midrule
        \multirow{3}{*}{PointNet} & FGSM \cite{sen2023adversarial}  & \multirow{3}{*}{SRS} & 68.9  & 9.89  & 4.83  & 66.5  & 3.6  & 20.5  & 67.6  & 2.29  & 25.4  
        \\
        & IFGSM \cite{you2023plant} &  & 75.3  & 4.8  & 38.5  & 70  & 4.8  & 23.5  & 80.1  & 1.6 &16.1    
        \\
         & PGD \cite{villegas2024evaluating} &  & 68.9  & 9.89  & 4.83   & 65.2  & 9.8  & 4.83  & 64.6 & 9.89  & \textbf{4.8}   
        \\
        \midrule
        \multirow{3}{*}{Dgcnn} & FGSM \cite{sen2023adversarial}  & \multirow{3}{*}{SRS} &  23.4 & 4.0  & 14.0  &  41.1   & 4.0  & 14.0  &  51.8 & 4.0  & 14.0  
        \\
        & IFGSM \cite{you2023plant} &  &34.7   & 5.7  & 19.2  & 70  & 5.7  & 19.2  & 68.6  & 5.7   & 19.2   
        \\
        & PGD \cite{villegas2024evaluating} &  &68.6   & 9.89  & 48.3  & 65.5  & 9.89  & 48.3  & 64.8  & 9.8  & 48.1 
        \\
        \midrule
        \multirow{3}{*}{pct} & FGSM \cite{sen2023adversarial} & \multirow{3}{*}{SRS} &  22 &3.9   & 13.9  & 42.3  & 3.9  & 13.9  &  48.7 &3.9   &13.9
        \\
        & IFGSM \cite{you2023plant} &   & 33.1  & 5.8  & 19.27  &  68.8 & 5.8  & 19.2  &  68.0 & 5.8  & 19.2
        \\
        & PGD \cite{villegas2024evaluating}&   & 69.5  & 9.8  & 48.3  & 65.5  & 9.7  & 47.2  & 64.9  & 9.7  & 47.2
        \\
        \midrule
        3D-Diffusion& (DifA)Ours & SRS & \textbf{85.0}  &  \textbf{1.5} & \textbf{6.7}  &  \textbf{82.3} & \textbf{1.5}  & \textbf{6.7}  & \textbf{81.2}  & \textbf{1.5}  & 6.7 
        \\
        \bottomrule
    \end{tabular}
    }
    % \vspace{0.5em}
    \label{tab:table3}
    % \vspace{-0.5em}
\end{table*}

\textbf{Black-box Performance And Comparison. } 
We comprehensively compare our 3D diffusion black-box attack with various baselines, including regular optimization-based attacks such as FGSM, IFGSM, and PGD. Specifically, our method is implemented with a reverse diffusion step size of 100 and $1t$ DCD optimization iterations, while all baselines adopt untargeted attacks with $\epsilon$ set to 0.32. The comparisons are conducted on the same RTX 3050 GPU, evaluating metrics such as attack success rate (ASR), Chamfer distance (CD), and Hausdorff distance (HD). The results listed in Table \ref{tab:table1} show that our method incurs the least geometric distance cost to achieve nearly 90\% ASR, with a lower time budget compared to regular optimization-based attacks. This aligns with our intuition that the diffusion model better preserves the point features from the original point clouds during adversarial example generation.

To evaluate the generalizability of these attack methods, we assessed the performance of adversarial point clouds after careful data preprocessing and measured the attack's success rate (ASR) across different models. The results demonstrate that our proposed diffusion attack consistently outperforms other methods in most cases in terms of ASR, CD, and HD, indicating its robustness and effectiveness across various 3D recognition models.

\subsection{Results on Defense Approaches - RQ2}
To further validate the robustness of each attack method, we perform performance evaluation tests on the produced adversarial examples after the defense methods. We consider common point cloud input preprocessing defense methods \cite{tang2023deep}. We demonstrate the superiority of our proposed scheme by testing it against other attack methods (FGSM, IFGSM, PGD). Input defense methods We choose two common input preprocessing schemes (SOR, SRS) for point clouds, The results are shown in Tables \ref{tab:table2} and \ref{tab:table3}. Moreover, the attack effect of our proposed method after SOR defense is almost the highest success rate among all attacks, and it also performs best in Chamfer Distance and Hausdorff Distance. 

\noindent \textbf{Performance comparison under SOR defense approach}
We use the SOR defense method with $k$ set to 2 and $\alpha$ using 1.1 as the defense parameter, and the results obtained show that our proposed scheme still has good aggressiveness after the point cloud defense treatment. In our experiments, we found that if the proxy model is Pointnet, the success rate in attacking PointNet++, Curvenet, and PointConv models is higher than if the proxy model is the other two. Using Dgcnn as the proxy model, the attack performance for Curvenet, PointConv is higher than that for PointNet++. This is an interesting phenomenon, suggesting that point cloud attacks can be ideally achieved by using similar models to generate adversarial examples for robustness tests. And the experimental results show that the attack method based on the diffusion generation model we proposed can produce better attack performance on different models through one-time proxy model generation, and the Chamfer Distance and Hausdorff Distance are smaller than other attack methods, indicating the superiority of our method. As shown in the figure, we show the visualization of adversarial point clouds of our proposed method and Shape-Invariant(SI) for comparison in Table \ref{tab:vision-compare}. The SI method is set to $\epsilon$ $=$ 0.16 and step size $=$ 0.007, The noise scale of our method is set to $0.05t$. Our adversarial generation scheme based on the diffusion model has fewer outliers and a smoother surface than the SI attack. Regardless of whether it has undergone the point cloud defense method, the point cloud quality generated by our adversarial sample generation scheme is superior to that of the SI attack.

% \jz{has fewer outliers? show statistics. Regardless of through the point cloud or not, it is not good wording. show through the point cloud compared with SI attack, what is the improvement and not through point cloud, what is the improvement}

% \begin{table}[t]
% \centering
% \scalebox{1.1}{
% \begin{tabular}{lll}
% \hspace{0.5cm}\textbf{Clean} & \hspace{0.7cm}\textbf{SI} & \hspace{0.5cm}\textbf{Ours} \\[0.02cm]
% \includegraphics[width=1.8cm]{image/sota compari/table.png} & 
% \includegraphics[width=1.8cm]{image/sota compari/SIatt.png} &
% \includegraphics[width=1.8cm]{image/sota compari/oursdiff.png} \\
% \includegraphics[width=1.9cm]{image/sota compari/chair.png} & 
% \includegraphics[width=1.9cm]{image/sota compari/SIsordefense-chair.png} & 
% \includegraphics[width=1.9cm]{image/sota compari/chair1.png} \\
% \end{tabular}}
% \caption{Experimental results comparison chart of our method and SI. Each column refers to a method. \textit{Upper row}: Illustration of an unprotected AEs. \textit{Lower row}: Illustration of AEs defended by SOR.}
% \label{tab:vision-compare}
% \end{table}

\textbf{Performance comparison under SRS defense approach:}
Although our proposed method shows good performance against SOR defense methods, it may be insufficient against some well-designed adversarial examples, for this reason, we conducted SRS defense experiments to improve the persuasiveness of our method.
For SRS defense method, we set the drop num points to 500, experimental results show that our proposed method has better attack performance and transferability in the face of both defense methods, and the distance shows that our attack does not significantly damage the original point cloud. Moreover, after the adversarial point cloud generated by other attack methods is passed through the SRS defense method, the attack success rate drops significantly. It can be seen that randomly discarding 500 points is effective in reducing adversarial losses. Overall our 3D diffusion attack has strong resistance and good transferability against these defenses.
\begin{table}[!htb]
    % \setlength{\belowcaptionskip}{1cm}
    % \scriptsize
	\footnotesize
    \centering
        \caption{Attack success rate (ASR) and Chamfer Distance of our method with PointNet++ and Curvenet on ShapeNet.}
    \setlength{\tabcolsep}{1mm}{
    \begin{tabular}{lcp{10mm}<{\centering}p{10mm}<{\centering}p{10mm}<{\centering}p{10mm}<{\centering}p{10mm}<{\centering}p{10mm}<{\centering}p{10mm}<{\centering}p{10mm}<{\centering}p{10mm}<{\centering}p{10mm}<{\centering}p{10mm}<{\centering}p{10mm}<{\centering}p{10mm}<{\centering}}
        \toprule
        & \multirow{3}{*}{Attack}
        & \multicolumn{2}{c}{PointNet++ \cite{qi2017pointnet++}} 
        & \multicolumn{2}{c}{Curvenet \cite{muzahid2020curvenet}}
        \\
        \cmidrule(lr){3-4}\cmidrule(lr){5-6}
        & &  ASR$\uparrow$ & CD$\downarrow$  & ASR$\uparrow$ & CD$\downarrow$ 
        \\
        & &  (\%) & ($10^{-2}$)& (\%) & ($10^{-2}$)  
        \\
        \midrule
         \multirow{3}{*}{Ours} & Chair  & 70  & 0.8  & 64  & 0.8 
        \\
        & Airplane  & 65  & 0.3  & 75  & 0.3    
        \\
         & Bench   & 82  & 1  & 93   & 1  
        \\
        \midrule
    \end{tabular}
    }
    % \vspace{0.5em}
    \label{tab:table4}
    % \vspace{-0.5em}
\end{table}

\subsection{Ablation Studies - RQ3}
The amount of noise added by the attack is a critical and configurable parameter that affects the performance of our method. We evaluated different numbers of DCD optimizations with a fixed number of diffusion model steps, and the results are shown in Fig \ref{fig:experiment_OP_ASR}. Our method is directly affected by the number of optimizations, and the performance of the attack progressively decreases as the number of optimizations decreases from 5$t$ to 1$t$.

\begin{table}[!htb]
    % \setlength{\belowcaptionskip}{1cm}
    % \scriptsize
	\footnotesize
    \centering
        \caption{Different loss function settings are used in the diffusion process and the impact of the order on the results.}
    \setlength{\tabcolsep}{1mm}{
    \begin{tabular}{lcp{10mm}<{\centering}p{10mm}<{\centering}p{10mm}<{\centering}p{10mm}<{\centering}p{10mm}<{\centering}p{10mm}<{\centering}p{10mm}<{\centering}p{10mm}<{\centering}p{10mm}<{\centering}p{10mm}<{\centering}p{10mm}<{\centering}p{10mm}<{\centering}p{10mm}<{\centering}}
        \toprule
        & \multirow{3}{*}{Loss}
        & \multicolumn{3}{c}{PointNet++ \cite{qi2017pointnet++}} 
        \\
        \cmidrule(lr){3-5}
        & &  ASR$\uparrow$ & CD$\downarrow$  & HD$\uparrow$ 
        \\
        \midrule
         \multirow{3}{*}{Ours} & CDC & 73  & 0.13  & 0.71  
        \\
        & MSE+CDC & 70& 0.06  & 0.46    
        \\
         & CDC+MSE   & 71  & 0.06  & 0.23 
        \\
        \midrule
    \end{tabular}
    }
    % \vspace{0.5em}
    \label{tab:table5}
    % \vspace{-0.5em}
\end{table}

In order to find the appropriate number of optimizations, we combine the success rate of the attack and the various metrics of the resistance to defense and the stealthiness of the generated adversarial examples, and use the 2t optimization as our experimental benchmark.

To investigate the impact of different loss functions during the diffusion process on the generation of adversarial samples, we conducted ablation experiments focusing on the sequence and combination of these loss functions. As shown in Table \ref{tab:table5}, the combination of DCD and MSE demonstrates a strong constraint on the noise added during the diffusion process while minimally affecting the attack success rate. These results suggest that the DCD+MSE combination effectively balances the imperceptibility of the perturbations and the overall attack performance. Due to time constraints, we only selected the Chair classification test set data for evaluation.

\begin{figure}[!htb]
\centering
\includegraphics[width=0.45\textwidth]{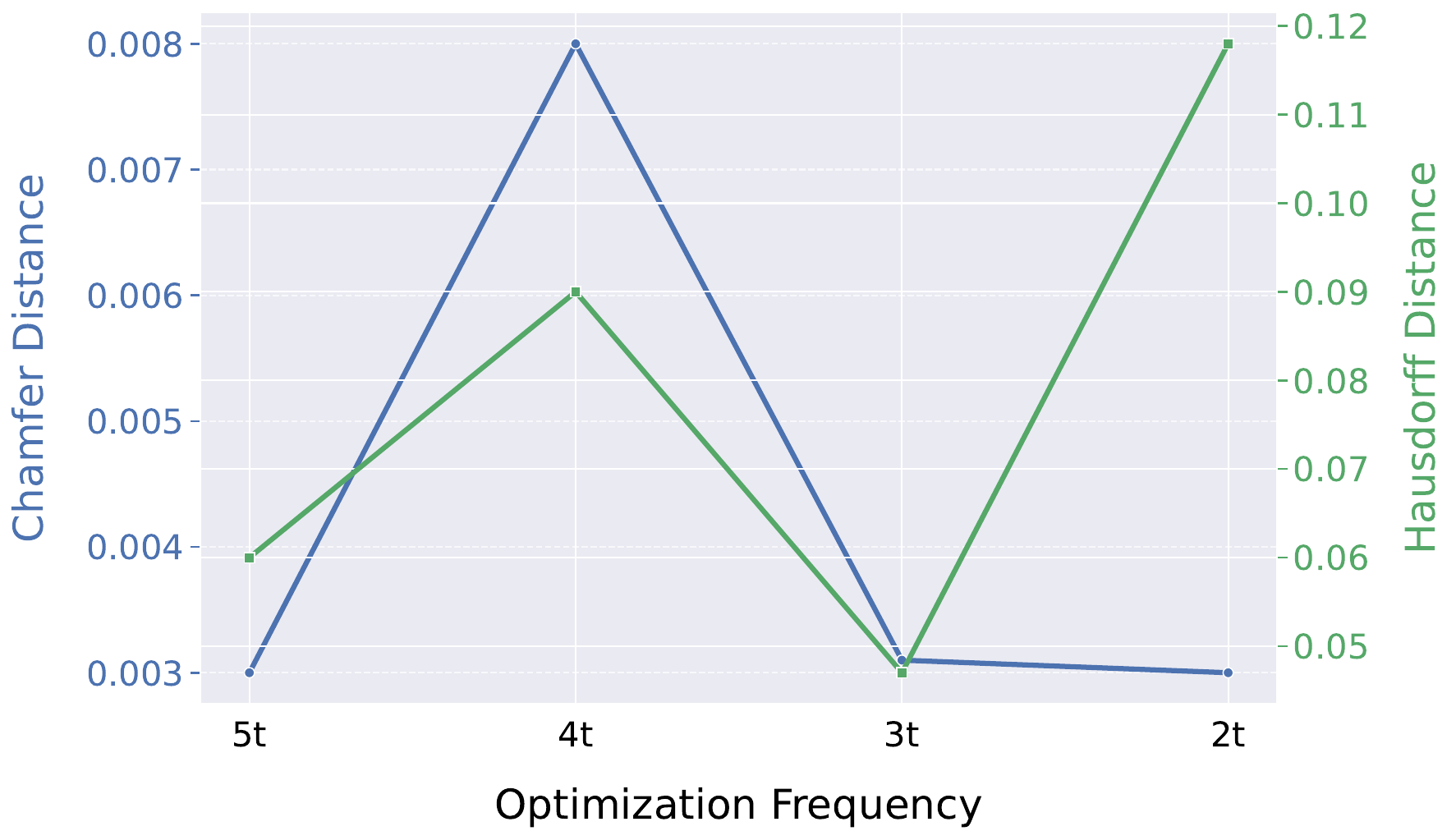}
\caption{Comparison experiments of Chamfer Distance and Hausdorff Distance under different settings of the number of DCD optimizations for diffusion attacks, with the attack target network as PointNet++.}
\label{fig:CD-HD}
\end{figure}

\section{Conclusion}
% A deceptive and imperceptible point cloud is successfully generated by 3D diffusion modeling to attack the target autopilot system. The attack is effective, causing the system to misjudge in the perception phase, which may lead to dangerous behaviors. The attack point cloud is highly imperceptible and difficult to detect by the naked eye, which increases the probability of a successful attack. The generated point cloud shows robustness and adaptability in the face of different scenes, lighting, and vehicle states. It successfully attacks the perception and decision-making modules of the target system. The attack model has a strong generalization ability, adapts to different versions of autopilot systems, and is adaptive to changes in the system.\jz{the conclusion is dry, what are the insights, limitations and future directions?}

The deceptive and imperceptible point cloud was successfully generated using 3D diffusion modeling to attack the target autopilot system. This attack proves to be effective, causing the system to misjudge in the perception phase, potentially leading to dangerous behaviors. The attack point cloud is imperceptible. The generated point cloud demonstrates robustness and adaptability across various scenes, lighting conditions, and vehicle states, effectively compromising the target system's perception and decision-making modules. Our proposed black-box adversarial sample generation method is capable of producing deceptive adversarial examples (AEs) with a certain degree of transferability. However, the method incurs significant time costs during execution. While generating black-box adversarial examples using generative models seems feasible, there remains room for improvement in reducing deformation compared to more advanced white-box approaches. In future work, we will continue to explore methods for generating black-box adversarial examples with less deformation from the perspective of the diffusion model encoder.

\bibliographystyle{ACM-Reference-Format}

% % --- Appendix ---%
% \appendix
% \input{appendix}
\end{document}